# Helical Bragg gratings: experimental verification of light orbital angular momentum conversion

A. G. Okhrimchuk, V. V. Likhov, S. A. Vasiliev, A. D. Pryamikov

*Abstract*—**Structured light, in particular light possessing orbital angular momentum (OAM), has been actively studied in recent decades. Helical Bragg grating (HBG) is a reflecting optical element, which predicted to be able to convert the OAM of light in waveguides and fibers. However, the HBG has not been demonstrated experimentally yet. Here we report the first experimental manifestation of HBG created in the form of a waveguide with depressed-index cladding. Few-mode channel helical-depressed-cladding waveguides (HDCWs) have been written in a YAG crystal by a femtosecond laser beam. The HDCW pitches were comparable to near-IR wavelengths. The spectral and polarization characteristics of the transmitted and reflected light have been studied. It was shown that HDCWs behave like HBGs providing narrowband resonance coupling of the counter propagating modes, the OAM of the modes differing by the Bragg order or, what is the same, the topological charge of the helical cladding structure. Adjusting the HDCW parameters allows one to excite a reflected wave at the predetermined wavelength and with the predetermined OAM change with respect to the light coupled to the waveguide. Thus, for the first time, the HBGs suitable for the generation and filtration of the vortex light have been experimentally demonstrated. The experimental results obtained agree well with the coupled mode theory calculations.**

*Index Terms*— optical waveguides, optical device fabrication, gratings, micromachining.

## I. INTRODUCTION

OPTICAL vortices with orbital angular momentum (OAM) discovered in 1989 [1], also known as beams with phase singularity, open up prospects for finding new effects of light interaction with matter. We have witnessed impressive achievements associated with OAM beams in atomic and electronic optics and also in the area of micro-scale matter manipulation[2], [3]. Light wave acquired a new dimension in its parameter space – OAM - along with wavevector, frequency, phase and polarization state. OAM beams are considered as promising for data transfer and processing in both free space and optical fibers [4]–[6]. The application of OAM beams is one of the mainstreams of modern optical communication. OAM beams also hold promise for increasing the information capacity of the protected communication channels based on quantum entanglement [7].

In bulk optics, beams with different OAM values are generated with the help of special holograms (fork-shaped, for example), helical phase plates, cylindrical lenses or space light modulators (SLM) [2]. In addition, OAM beams can be obtained by special optical elements, such as Q-plates [8] (including those written by direct laser writing (DLW)[9]) and metasurfaces [10].

In optical fibers and cylindrical channel waveguides, the set of optical vortex modes (OAM modes) is regarded as a separate mode basis similar to vector mode or LP mode bases[11]. Excitation of OAM modes is mostly performed by relatively narrowband (~10 nm) mode coupling between the initial fundamental mode and modes with a higher azimuthal order, the latter being equal to OAM of vortex mode. This takes place in optical fiber couplers, where two neighboring fiber cores exchange energy [12], [13], or in fiber gratings, where the required phase matching is provided by periodic perturbation[14]–[17]. A wider OAM-mode excitation band (>100 nm) can be achieved by means of photonic lanterns [18], [19].

Periodic structures in optical fibers and channel waveguides can be created by alternating the refractive index or geometry of the waveguide. In both cases, spectrally selective and efficient (90% and higher) mode coupling with a change in azimuthal order can take place. Nowadays periodic structures with a relatively large period (hundreds of micrometers) coupling copropagating modes (long-period grating, LPG), are used to excite OAM modes in fibers. The LPG spectral selectivity is relatively poor (~10 nm), whereas fiber Bragg gratings (FBGs) can couple counter-propagating fiber modes with much better selectivity (~0.1 nm). High spectral selectivity is an intrinsic feature of HBGs as well. Moreover, HBGs possess a topological charge and therefore can change OAM of light[20], [21]. Meanwhile, despite considerable research and practical interest, as far as we know, such helical gratings with a pitch comparable to the wavelength order have not been realized experimentally, although some methods for their fabrication have been proposed[22], [23]. The problem is

A. G. Okhrimchuk (okhrim@fo.gpi.ru) and V. V. Likhov (vladislavlikhov@gmail.com) are with Prokhorov General Physics Institute of Russian Academy of Sciences, Dianov Fiber Optics Research Center, 38 Vavilov Str, Moscow 119333, Russia, and with Mendeleev University of

Chemical Technology of Russia, 9 Miusskaya Square, Moscow 125047, Russia.

S. V. Vasiliev (sav@fo.gpi.ru) and A. D. Pryamikov (pryamikov@fo.gpi.ru) are with Prokhorov General Physics Institute of Russian Academy of Sciences, Dianov Fiber Optics Research Center, 38 Vavilov Str, Moscow 119333.

technically complicated and cannot be solved in the frame of traditional FBG writing or helical fiber drawing techniques.

Direct laser writing (DLW) by ultrashort infrared pulses is a unique micromachining method as it allows flexible 3D structuring of transparent materials with a submicron spatial resolution. This technique makes it possible to create waveguiding structures of different types inside transparent dielectrics [24]–[26]. The depressed cladding structure allows creating a tubular waveguide and was originally employed in crystals [27], [28]. Waveguides of this type is usually inscribed in crystals as cavities of micro-lasers operating in continuous-wave or pulsed regimes [29]–[33]. Multimode helical depressed cladding waveguide (HDCW) written in a YAG crystal by means of a femtosecond (fs) laser beam has been described and applied for supercontinuum generation in visible [34], [35]. Then we preliminary investigated the HDCW reflection properties and have found that Gaussian is converted to OAM modes under resonance Bragg reflection [36], [37].

Therefore, this work is dedicated to a detailed study of the transmittance and reflectance of DLW-written HDCWs with their helical pitch optimized for Bragg reflection in the 1.55 μm wavelength range. It has been experimentally shown that the HDCWs manifest the spectral properties of HBGs, the OAM of the reflected light being changed by the Bragg diffraction order (the topological charge of the helical structure).

## II. EXPERIMENTAL METHODS AND RESULTS

### A. Inscription of HDCW

A femtosecond laser beam with a wavelength of 1030 nm, pulse duration of 180 fs, repetition rate of 3 or 15 kHz and pulse energy of 400 nJ was used for the writing of the helical depressed cladding waveguides (HDCW). The laser beam was focused in the volume of Nd:YAG (0.8 at.%) by a 50x Olympus objective lens (NA=0.65) equipped with a correction collar set for the writing depth of ~200 μm. A concave cylindrical lens was placed in front of the objective lens to introduce astigmatism into the beam and, therefore, to form two beam waists of lenticular shape of diameter about 10 μm, the closest to the objective lens being the writing one.

It is known that DLW in YAG crystal under low repetition rate forms regions of depressed refractive index[28], [29]. Therefore, the channel waveguides can be formed by modification of the refractive index in the tubular region around pristine material of the waveguide core [27]–[29], [31], [33], [38].

The crystal was a cuboid with dimensions of 2.3 x 5.3 x 9.65 mm. All crystal facets were polished with laser quality. Optical axes of written waveguides (Z-axis in Fig. 1) were parallel to [111] crystallographic axis directed along the longest edge of the crystal sample. The laser beam propagated along Y axis was focused through the facet parallel to (211) crystallographic plane. The sample was placed on a high-precision 3D linear translation stage Aerotech ABL1000 and scanned with a constant velocity along the Z-axis relative to the lenticular beam waist, flattened in the same Z-direction. The sample was simultaneously moved along a counterclockwise elliptical path in the X-Y plane with the ellipse diameters $D_x = D_{sp}$ and $D_y = D_{sp}/n_0$, where $D_{sp}$ is the designed circular ring diameter (Fig.1), $n_0$ is the YAG crystal refractive index at 1030 nm. HDCW cladding with a constant radial thickness was inscribed by this technique[34].

The cladding of waveguide was inscribed by femtosecond pulses, the laser beam waist being moved along the helical trajectory as it was done in [38]. The principal difference of our approach is application of the beam of the lenticular shaped beam waist in contrast to the traditional cigar-like beam waist used in [38]. The lenticular-shaped beam waist provides the way to create a thick and azimuthally uniform depressed-index cladding with less mode leakage[38]. These advantages allowed us to inscribe helical cladding structures with expanded pitch, thereby ensuring significant index modulation and, consequently, the efficient Bragg reflection.

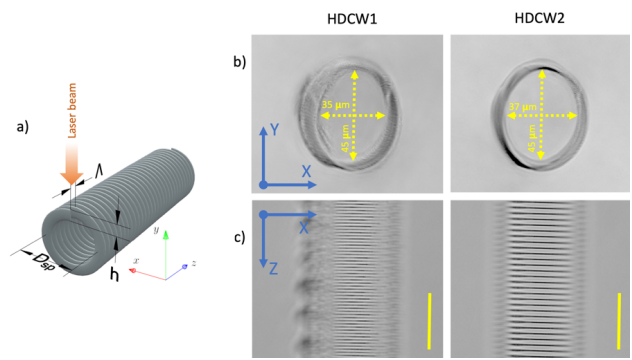

Fig.1. HDCW scheme (a), $\Lambda$ is pitch of helix, $D_{sp}$ is diameter of the helical movement during laser writing, $h$ is radial thickness of HDCW. Bright-field-microscopy images of HDCWs written in YAG:Nd single crystal end view (b) and side view (c) taken under backside illumination. Helix pitches equal to 1.7 and 3.0 μm for HDCW1 and HDCW2 respectively. HDCW1 and HDCW2 were inscribed with the same pulse energy and duration, at 15 and 3 kHz repetition rates, respectively. A 30 μm scale is included.

The frequency of the elliptical motion of the 3D translation stage was $f_{osc}$=15 Hz. The scan velocity along the Z-axis was set to obtain the required helix pitch $\Lambda$: $V=f_{osc}\Lambda$. Two waveguides with pitches of 1.7 μm (HDCW1) and 3 μm (HDCW2) were written and investigated (Fig. 1b). The pulse repetition rate was 15 kHz for the HDCW1 and 3 kHz for the HDCW2, whereas the other parameters of laser pulses were similar. In both cases HDCW cross-section proved not to be round, its slight ellipticity is explained by uncompensated spherical aberrations.

The chosen pitches correspond to Bragg diffraction orders of $p$=4 and $p$=7 around wavelength of 1.55 μm for the HDCW1 and HDCW2, respectively. The total length of each waveguide was 9 mm, but the 3D translation stage did not ensure the required precision of the pitch along the entire length of the waveguide. As follows from the Fourier analysis of the lateral microphotographs of the waveguides, the fragment of which is shown in Fig. 1c, the pitch of the helix undergoes two 0.5% hops along the waveguide length. As this took place, the waveguide parts between the hops were uniform with pitch variations as small as 0.01%. Thus, each HDCW consisted of three uniform parts, several millimeters in length, each of which was an individual HBG with its own resonance wavelength different from the neighboring one by ~8 nm. The results reported here correspond to the parts adjacent to the input ends of the HDCWs. The parts were 2.1-mm and 3.9-mm long for HDCW1 and HDCW2 respectively.

To estimate the refractive index depression in the fs-modified regions of the crystal, straight linear tracks were written inside an identical crystal with the same pulse energy, repetition rate and linear translation velocity. This resulted in the spatial pulse overlapping the same as that in the helices. The refractive index depression measured with microscope equipped with QPM [39] at the wavelength of 503 nm proved to be 0.015 for the HDCW1 and 0.01 for the HDCW2. The measurements were performed using a microscope objective with NA=0.3.

### B. Characterization of the HDCWs

Fig. 2 shows the scheme of the optical setup used for HDCW characterization. Transmittance and reflectance spectra, as well as the mode profiles were measured at wavelengths near 1.55 μm. Single-frequency laser diode tunable in the range of 1480 – 1620 nm (Thorlabs TLK-L1550, Littman configuration, the linewidth is approximately 100 kHz) equipped with an SMF-28 fiber pigtail was used as the light source. The laser light was collimated by lens L1 into a Gaussian-like beam with a diameter of ~2 mm. Linear horizontal polarization in front of beam splitter BS1 was adjusted by polarization controller PC. Hereafter, polarization state of light coupled into the waveguide was controlled by a half- or quarter-wave plate WP, which provided linear vertical or circular polarization states, respectively. A laser beam was coupled into the waveguide by means of two lenses L2 and L3, the focal lengths being 250 and 20 mm, respectively. By controlling the distance between the lenses, it was possible to finely adjust the beam waist size at the input of the waveguide and, therefore, to optimize the coupling efficiency to obtain the maximum transmittance. Then the maximum reflectance was obtained by adjusting the axial position of lens L3 only, which resulted in increasing the distance between the crystal input end and the beam waist by 1-2 mm, whereas the beam waist itself remained unchanged. The L3 adjusting for better transmittance was performed beyond the Bragg resonance, whereas the reflectance was maximazed within the reflection bands (1536.5 nm for the HDCW1 and 1549.0 nm for the HDCW2). The transmission properties of the waveguides presented below were measured in the former L3 position whereas the reflection properties in latter one, unless otherwise stated.

Both input and output ends of the crystal were antireflection coated. The crystal with the written HDCWs was placed on a high-precision 6-axis stage Thorlabs MAX601.

The HDCW output light was analyzed in near field with a 10x objective L4 which imaged the output end of the waveguide on an InGaAs sensor of the ArtCam-0016TNIR-

1 camera (C1). For far field measurements, a 10x objective lens was replaced by an M3514-SW objective (f=35 mm, Computar) placed in front of the camera so that the camera matrix was in the focal plane of the objective.

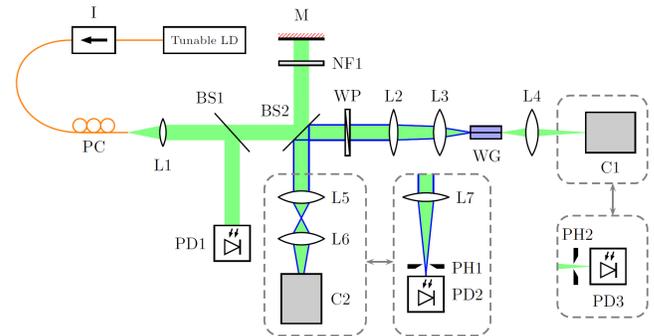

Fig.2. Scheme of the optical setup for the HDCW characterization. I is an optical isolator, PC is a polarization controller with an SMF-28 optical fiber pigtail, BS1, BS2 are beam splitters, WP is a half- or quarter-wave plate, M is a metallic mirror, NF1 is a neutral-density filter, PD1, PD2 are photodiodes, PH1, PH2 are iris diaphragms, C1, C2 are IR cameras. Focal lengths of the lenses are 12 mm (L1), 250 mm (L2), 20 mm (L3), , 35 mm (L5), 35 mm (L6, objective for far-field measurements), 700 mm (L7). L4 is the 10x objective. Light reflected by the HDCW is shown by blue lines.

Fig.3 shows the intensity profiles measured at the HDCW1 and HDCW2 outputs in the near and far fields. As seen in Fig.3(a,b) the waveguides effectively confine the coupled light . The intensity profiles in the near field are similar to the Gaussian shape. The far-field profiles have a more complicated structure consisting of several axially symmetrical rings (Fig.3(c,d)). These patterns point to the excitation of not only the fundamental waveguide mode LP01, but also higher-order modes. The interference of the fundamental and higher-order modes is likely to lead to the ring-shaped intensity patterns in the far field of the light passed through the waveguides [40].

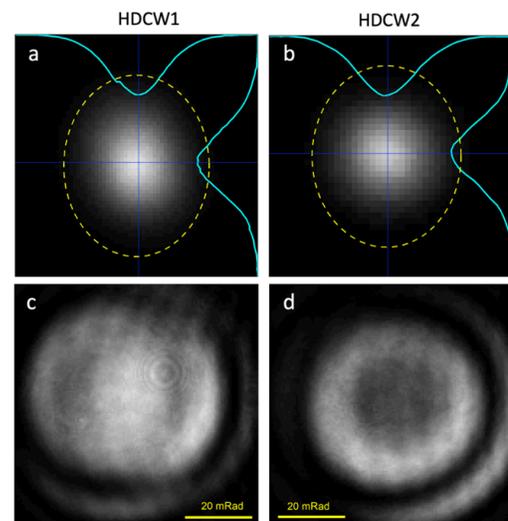

Fig.3. Intensity profiles of the output beams of HDCW1 (a,c) and HDCW2 (b,d) measured in the near field (a,b) and in the far field (c,d) under the conditions of 1560-nm linear-polarized laser beam excitation. Waveguide core boundaries are shown by yellow dashed lines, whereas the cross sections of the light intensity distributions are given by cyan solid lines (a,b).

Photodiode PD2 and PD3 were used to measure the HDCW reflection and transmission spectra respectively, iris diaphragms PH1 and PH2 were installed to filter the light coming from the waveguide core. Photodiode PD1 was used to take into account variation of the laser power. To evaluate the HDCW reflection coefficient, the reflection signal from the waveguide was compared with a Fresnel reflection from an uncoated YAG crystal placed instead of the sample.

Fig.4 shows the transmittance and reflectance spectra of the waveguides measured at different polarization states of the input light. The sharp jumps in the spectra, caused by mode hopping in the tunable laser diode, only slightly distort the shape of the spectra. It turned out that the transmittance and reflectance bands are complementary to each other, and the spectra are similar for all polarization states of the coupled light. A small difference among the band amplitudes is obviously explained by the polarization sensitivity of beam splitter BS2 (Fig.2).

splitter BS2 and lenses L2, L3 and L5. Objective L6 displayed the far field pattern on camera C2. The far field structure was investigated by means of an interferometric method as follows: the reference Gaussian beam reflected by mirror M was superimposed on the investigated beam reflected from the waveguide, with the reference beam intensity being adjusted by neutral-density filter NF1.

The structure of the far fields obtained for HDCW1 is intricate as its reflection band is composite (Fig.5 (a,b,c)). The fields are ring-shaped, which is typical for vortex waves with phase singularity in the center. The interference pattern of the reflected light with the coherent reference Gaussian beam consisted, as a rule, of four branches (OAM = 4, Fig.5(e)). Nevertheless, some deviations indicating the presence of vortex fields with larger or smaller orbital angular momenta (OAM=3 in Fig.5(f), and OAM=5 in Fig.5(d)) are observed at the short and long wavelength parts of the spectrum respectively.

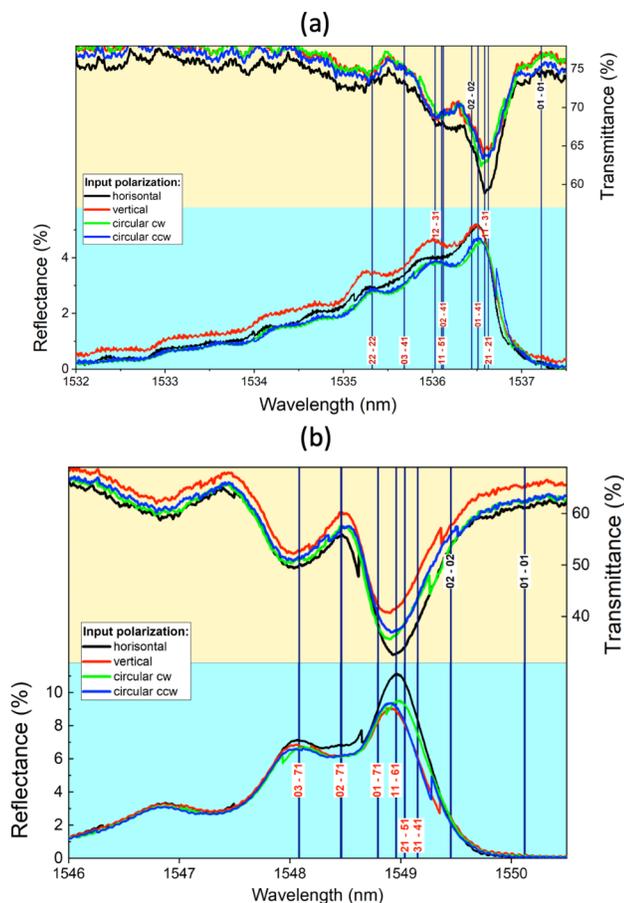

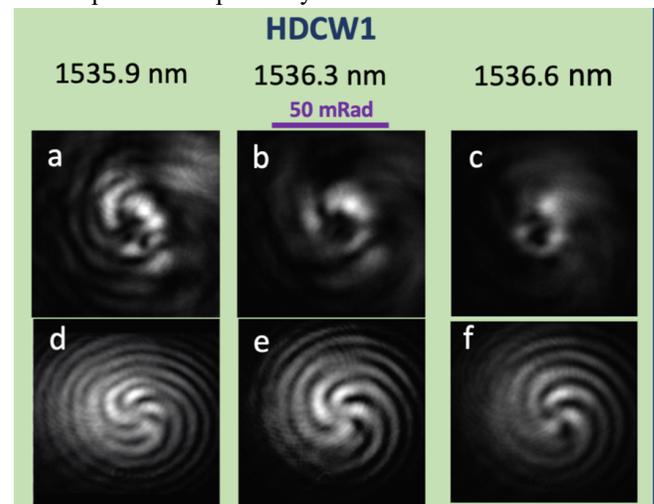

Fig. 5. a-c) Far field intensity profiles of the modes reflected by HDCW1 at different wavelengths. d-f) The corresponding far field interference patterns of the reflected light with the reference beam.

A band centered at 1549 nm in the HDCW2 reflectance spectrum is the most pronounced (Fig. 4(b)). The far field patterns of the light wthin this band consist of concentric rings as in the case of HDCW1 (Fig.6 (c,d)). In contrast to HDCW1 the interference patterns observed for HDCW2 were clear for both types of the input beam couplings (Fig6 (e-h)). Namely, the patterns with six branches were observed under both couplings at the band center and in the longer wavelength edge of the band. Sharp seven branches were observed at the short wavelength edge of this band (1548.5 nm in Fig.6) at the input coupling was optimized for the reflection, whereas more complicated interference pattern was observed at the coupling optimized for transmission.

The interference patterns shown in Fig.5 and Fig.6 were registered at linear polarization states of the probe light and were independent from the polarization direction.

Supplementary videos of the interference patterns recorded while increasing the laser wavelength through the reflectance bands will be available at **http://ieeexplore.ieee.org**.

Fig.4. The transmittance and reflectance spectra of HDCW1 (a) and HDCW2 (b) measured for the beams with the linear and circular polarizations of light coupled to the waveguides (polarization states are indicated in the Figures). The calculated Bragg resonance wavelengths are shown by the vertical solid lines labeled by the coupling mode indexes: the first- and second-digit pairs are indexes of the forward and backward modes respectively. The mode pairs satisfying both matching conditions Eq.4 and Eq.5 are shown in red, whereas the mode pairs satisfying only the phase matching condition Eq.4 are shown in black.

The light reflected by the HDCW was investigated in the far field at several wavelengths either with, or without the reference beam reflected by metallic mirror M (Fig.2). The image of the HDCW input end was formed by means of beam

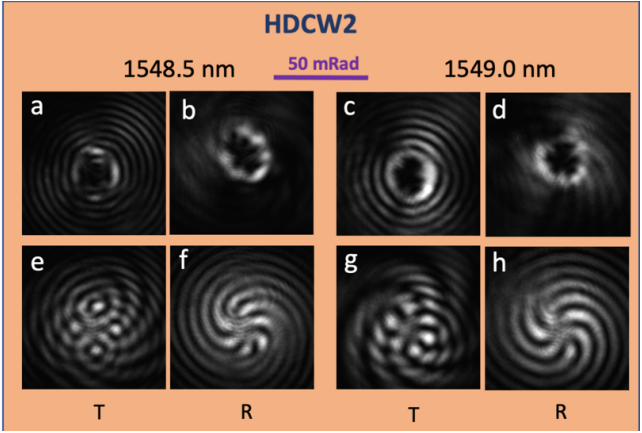

Fig.6. Far field intensity profiles of the modes reflected by HDCW2 at 1548.5 nm (a, b) and 1549.0 nm (c, d). The corresponding far field interference patterns of the reflected light with the reference light at 1548.5 nm (e, f) and 1549.0 nm (g, h). Input coupling is optimized for transmission (a,c,e,g) or reflection (b,d,f,h) maximum and marked by T and R respectively.

## III. DISCUSSIONS

In the general case, the spatial distribution of the refractive index in HBG can be described by Fourier series in cylindrical coordinate system $(r, \varphi, z)$:

$$n(r,\varphi,z) = n_0 + \Delta n(r,\varphi,z)$$
$$= n_0 + \delta n(r)\sum_{p=0}^{+\infty} A_p \cos\left[p\left(\alpha\varphi - Kz + \theta_p\right)\right], \quad (1)$$

where $n_0$ is the refractive index of an unmodified crystal, $\delta n(r)$ is the radial distribution of the index change produced by DLW ($\delta n(r)$ is nonzero inside the helix cladding only). The first term of the sum ($p=0$) describes the average index change, and the rest terms are responsible for the index modulation with corresponding space and angular frequencies. In Fourier series (1) the argument is the helicoidal coordinate $\xi = \alpha\varphi - Kz$ as shown the supplementary Appendix. Such a representation allows one to describe the periodic index variation in the same way both in angular $\varphi$ and axial $z$ coordinates for waveguide cladding consisting of one or many embedded helixes with equal pitches $\Lambda$ ($K=2\pi/\Lambda$ is the helical constant). $A_p$ и $\theta_p$ are the amplitude and phase of the p harmonic, handedness $\alpha=+1$ for the right-handed, and $\alpha=-1$ left- handed helixes.

When a helical index structure is considered, it is convenient to express the eigen solutions of the wave equation in terms of the vortex modes (a mode basis with different OAMs). Thus, the transverse electrical field in the considered waveguide can be expressed as a linear combination of the transverse electric fields of the OAM modes [2], [41]:

$$\vec{E}_t(r,\varphi,z) = \frac{(\vec{x} + s\vec{y})}{\sqrt{1+\left|s\right|^2}} \sum_{q=-\infty}^{+\infty}\sum_{m=1}^{+\infty} C_{qm} F_{qm}(r) e^{i(q\varphi - \beta_{qm}z)}, \quad (2)$$

where s is a complex parameter defining the OAM-mode polarization state, $C_{qm}$ is the mode amplitude ($q$ is the azimuthal index or OAM, and $m$ is radial mode indices),

$F_{qm}(r)$ is the normalized radial profile of the OAM transverse field components, $\beta_{qm}$ is the propagation constant.

Resonance coupling arises between counterpropagating modes with the same polarization states if the helix pitch is comparable with the light wavelength. According to the coupled mode theory [42] the specific mode coupling coefficient g can be calculated:

$$g = \frac{2n_0\omega\varepsilon_0}{P}\int_0^\infty\int_0^{2\pi}\vec{E}_t^{*,forw}\Delta n(r,\varphi,z)\vec{E}_t^{back}r\,dr\,d\varphi, \quad (3)$$

where $P$ is the power of the forward propagating mode. Analysis of (3) gives that efficient coupling takes place if both phase matching (4) and angular momentum (5) matching conditions are satisfied simultaneously.

$$\beta_{qm} + \beta_{lk} = pK, \quad (4)$$

$$q - l = \alpha p, \quad (5)$$

where $q$ and $l$ are OAMs of forward and backward propagating waves correspondingly, and $m$ and $k$ are radial mode indexes. Equations (4) and (5) are similar to phase matching and angular momentum conditions obtained for long-period gratings [43]. According to these conditions, OAMs of forward and reflected light differ by the helix topological charge, which is equal to the Bragg order of the HBG $p$. In addition, corresponding coefficient in Fourier series should be non-zero (see Appendix). It should be noted that projection of OAM vectors $q$ and $l$ are considered in the laboratory coordinate system (Z axis is directed along the forward propagating wave), therefore sign of $l$ is opposite if the reflected light is considered in the coordinate system with Z axis directed along the wave propagation.

The observed spiral interference patterns of the light reflected by the HDCW testify that we are dealing with optical vortices [2], [4], [44]. In the experiments with HDCW1, the patterns observed at different wavelengths mean that the reflected light carries OAM equal to three, four or five ($l=3,4,5$, Fig.5(d-f)), whereas HDCW2 produced vortex light with OAM equal to six or seven ($l=6,7$, Fig.6(e-h)). In accordance with the Bragg phase matching condition (Eq.4), the fourth reflection order ($p=4$) for HDCW1 and the seventh reflection order ($p=7$) for HDCW2 occur in 1.5-μm wavelength range. Comparison of (4) with the angular momentum matching condition (5) clearly shows that the $p$-values are also the topological charges of our waveguides. Therefore, it is sufficient to excite a copropagating mode without angular momentum ($q = 0$, i.e. $LP_{0m}$ mode) to reflect an OAM mode with $l=4$ in HDCW1 (Fig.5e) and with $l = 7$ in HDCW2 (Fig.6h). At the same time, for reflection of modes with $l=3$ (Fig.5f) and $l =5$ (Fig.5d) in HDCW1, and $l=6$ (Fig.6f) in HDCW2, modes with non-zero OAM, i.e. with $|q| = 1$ ($LP_{1m}$ mode) must be initially excited. We think that when a Gaussian beam is coupled to the waveguides, pairs of OAM modes with opposite OAM signs ($q=\pm1$) are also excited together with the fundamental mode with zero OAM. This could result from the well-known linear combination of even and odd $LP_{1m}$ modes:

$OAM_{\pm 1} = LP_{1m}^{even} \pm iLP_{1m}^{odd}$ [11]. The latter assumption makes it possible to explain the experimentally observed interference patterns with a different number of branches taking into account the angular momentum conservation law.

The complex propagation constants of the HDCW modes were numerically calculated in the $LP$-mode approximation using COMSOL Multiphysics software. This approach allowed us to find the resonance wavelengths for mode coupling satisfying (4) as well as the propagation losses of the modes. We supposed that the input Gaussian laser beam can be decomposed in a set of low-order modes with $|q|<2$. To calculate the waveguiding properties of the HDCWs, we neglected the helical shape of the cladding and substituted it with a solid depressed-index tubular region. The overall geometry of the model tube (transverse dimensions and wall thickness) was equivalent to the laser-inscribed cladding of the HDCWs (see Fig.1), the index difference being taken as $3.5 \cdot 10^{-3}$ and $1.3 \cdot 10^{-3}$ for HDCW1 and HDCW2, respectively.

The calculated resonance wavelengths are shown in Fig.4 by vertical lines. As seen in Fig.4(a), the most intense reflection bands in the HDCW1 spectrum ($p=4$) can be due to the following mode coupling: $LP_{01}$—$LP_{41}$, $LP_{02}$—$LP_{41}$, $LP_{11}$—$LP_{31}$, $LP_{12}$—$LP_{31}$, and $LP_{11}$—$LP_{31}$. $LP_{22}$ mode self-coupling can contribute to the short wavelength edge of the spectrum (the $LP_{22}$—$LP_{22}$ resonance is just a conversion of the counter-propagating OAM modes with $q=-2$ and $l=+2$). The four-branch interference pattern prevailing in the range of 1535.9-1536.6 nm confirms the major contribution of $LP_{01}$—$LP_{41}$ mode coupling (Fig.5(e) and Supplementary Video_1), whereas the appearance of irregular patterns with three ($\lambda>1536.6$ nm) and five ($\lambda<1535.9$ nm) branches points to small contribution of $LP_{11}$—$LP_{31}$ and $LP_{11}$—$LP_{51}$ mode coupling, respectively (Fig.5(f), Fig.5(d) and Supplementary Video_1). Finally, it is possible that $LP_{21}$—$LP_{21}$ mode coupling (Fig.4(a)) manifests itself by some disintegration of the interference pattern at certain wavelengths within the 1536-1537 nm spectral range (supplementary Video_1). The calculated propagation losses for all the modes considered for HDCW1 did not exceed 0.3 cm$^{-1}$.

The most intense band in the reflection HDCW2 spectrum ($p=7$) at 1549 nm is mainly due to the $LP_{11}$—$LP_{61}$ mode coupling, which leads to the excitation of the OAM mode with $l=6$ for both types of the input beam coupling. The interference patterns with six branches visible in range of 1548.9 – 1549.1 nm correspond to this Bragg condition ((Fig.6(g,h) and supplementary Video_2). The less intense reflection in the range of 1548.1 - 1548.8 nm is due to the $LP_{03}$—$LP_{71}$ mode coupling, which leads to the reflection of the beam only with $l=7$ at the coupling optimized for reflection (Fig. 6(f)) and Supplementary Video_2). In this case the fundamental mode ($q=0$) was predominantly excited, therefore only $LP_{71}$ mode is back reflected. A competition of reflected beams with $l=6$ and $l=7$ in the interference pattern (Fig.6(e)) have been observed at the coupling optimized for transmission. The complex but a regular pattern of this interference is apparently due to the difference in the positions of the centers of the $LP_{61}$ and $LP_{71}$ modes caused by the waveguide ellipticity.

The domination in the reflectance spectrum of the band centered at 1549 nm corresponding to the $LP_{11}$—$LP_{61}$ mode coupling can be also explained by lower losses of the $LP_{61}$ mode ($\sim 10$ cm$^{-1}$ according to our calculations) with respect to the $LP_{71}$ mode ($\sim 20$ cm$^{-1}$). Nevertheless, the $LP_{01}$—$LP_{71}$ mode coupling contributes significantly to the complementary intense band in the transmission spectrum. This fact could also explain a small shift of the transmission band center to shorter wavelength with respect to the reflection band center (Fig.4(b)).

No conventional Bragg reflection (for example, the $LP_{01}$—$LP_{01}$ mode coupling) was observed for both HDCWs as can be seen in Fig.4 (hypothetical resonances are labeled by black indices). This fact is not surprising because the angular momentum matching condition (8) is not satisfied in this case.

The intensity profile of an OAM mode must have a ring shape. Indeed, the ring-like intensity patterns were observed in our experiments with a blocked reference beam (Fig.5 and Fig.6). Nevertheless, the purity of the patterns is significantly disturbed by the interference of two or even more reflected modes with different OAM values. Better ring purity was observed for the light reflected by HDCW2 because the reflected modes of different orders were better separated spectrally than in HDCW1. The interference of densely located resonances in HDCW1 ($LP_{01}$—$LP_{41}$, $LP_{02}$—$LP_{41}$, $LP_{11}$—$LP_{31}$, $LP_{12}$—$LP_{31}$, and $LP_{11}$—$LP_{51}$) leads to substantial perturbation of the ring intensity pattern (Fig. 4a-c). The pattern purity violation was most pronounced when the Gaussian beam coupling to HDCW1 was optimized for the transmission maximum, as in this case more modes are reflected (the corresponding figures are not shown for the sake of brevity), and therefore no clear interference patterns were observed in contrast to HDCW2. HDCW2 was found to be less sensitive to the type of input coupling, which is naturally explained by larger difference between propagation constants of neighboring modes.

The reflection coefficients at the band maxima for both waveguides amounted to 5% and 11% for HDCW1 and HDCW2, respectively, if excitation with horizontal polarization shown in Fig.4 is considered. However, the dips in the corresponding transmission spectra were a few times larger (20% and 50% for HDCW1 and HDCW2, respectively). This discrepancy can be explained by significant propagation losses of $LP_{41}$, $LP_{61}$ and $LP_{71}$ modes, which could be even higher than the calculated leakage loss due to scattering on the waveguide imperfections (> 5 cm$^{-1}$). The spectral widths of the HBG resonances for HDCW1 and HDCW2 were measured to be 0.35 nm and 0.6 nm, respectively. These values allowed us to estimate the effective HBG lengths, which proved to be 1.7 mm and 1.0 mm for HDCW1 and HDCW2, respectively. The estimated HBG length for HDCW1 is only slightly shorter than the length of the first part of this waveguide (2.1 mm), whereas the length of HBG in HDCW2 turned out to be substantially shorter than the length of the first part of HDCW2 (3.9 mm). This difference could be also explained by the large leakage (propagation loss > 10 cm$^{-1}$) for modes with high azimuthal orders ($LP_{61}$ and $LP_{71}$). Thus, the reflection coefficients of both waveguides are not limited by HBG lengths, but by the propagation losses of reflected high order OAM modes.

The known HBG length and transmittance at the resonance wavelength allowed estimating the lower value of the coupling coefficient $g$ for each HBGs. In spite of the rather large Bragg diffraction orders, the coupling coefficients for both waveguides were found to be greater than 2 cm$^{-1}$.

We believe that further optimization of the waveguiding structure will improve the quality of the HBG reflection, i.e. efficiency and purity of the reflected bands. Particularly, a few modes ring-core waveguide with DLW-inscribed HBG could be considered. A helical ring-core waveguide inscribed by solely DLW technique is also promising for this purpose. This waveguide type provides the increased mode overlapping together with the larger difference between propagation constants of different OAM.

## IV. Conclusions`

In summary, we have established and experimentally verified the helical Bragg grating technology. The gratings were fabricated in a YAG single crystal by direct femtosecond laser writing as helical depressed cladding waveguides. The beam waist of the lenticular shape was utilized to ensure large and uniform spiral cladding thickness regardless of the azimuth angle. It was shown that the fabricated HBGs efficiently excite light waves having an orbital angular momentum (optical vortices). Resonance mode coupling (Bragg reflection) in the wavelength range of 1.55 μm in the 4th and 7th Bragg diffraction orders was experimentally demonstrated for waveguides with 1.7 μm and 3 μm helical pitch, respectively. The gratings alter the OAM of the input mode by the value of Bragg diffraction order (or by the topological charge of the helical structure). We proposed to describe an arbitrary helical Bragg grating structure with a given period by a Fourier series with a helical coordinate as the argument, each term of the series corresponding to a certain Bragg diffraction order. It was shown that the topological charge of the grating at a certain resonance wavelength is equal to the diffraction order. The positions of the bands in the HDCW reflection and transmission spectra were calculated in terms of $LP$ modes which approximated the real OAM eigenmodes. The calculations explained most of the experimental spectral characteristics and the interference patterns occurring in such helical structures.

We suppose that the approach we suggested and described here paves the way for creation of narrowband optical elements suitable for OAM light generation and conversion.